\documentclass{article}
\usepackage{amsfonts}
\title{Formal similarity between mathematical structures of electrodynamics and quantum mechanics}
\author{A. A. Deriglazov\footnote{alexei.deriglazov@ufjf.edu.br ~ On leave of
absence from Dept. Math. Phys., Tomsk Polytechnical University,
Tomsk, Russia.}}
\date{Dept. de Matematica, ICE, Universidade Federal de Juiz de Fora,\\
MG, Brazil}
\begin{document}
\maketitle
\large

\begin{abstract}
Electromagnetic phenomena can be described by Maxwell equations written for the vectors of electric and magnetic
field. Equivalently, electrodynamics can be reformulated in terms of an electromagnetic vector potential.
We demonstrate that the Schr\"odinger equation admits an analogous treatment. We present a Lagrangian theory of
a real scalar
field $\phi$ whose equation of motion turns out to be equivalent to the Schr\"odinger equation with time independent potential. After introduction
the field into the formalism, its mathematical structure becomes analogous to those of electrodynamics.
The field $\phi$
is in the same relation to the real and imaginary part of a wave function as the vector
potential is in respect to electric and magnetic fields. Preservation of quantum-mechanics probability is just an energy conservation law of the field $\phi$.
\end{abstract}

\noindent
\noindent
\section{Introduction}
Lagrangian formalism is based on Euler-Lagrange equations of motion. They represent
a system of second-order differential equations written for a set of variables that describe the
position of a physical system of interest. Hamiltonian formulation suggests an equivalent description in terms
of first-order equations written for independent variables describing the position and velocity of the system.

From the beginning, some important equations of physics have been formulated in the
Hamiltonian-type form.
In particular, a pair of Maxwell equations, containing the temporal derivative, represents a Hamiltonian system.

For a mechanical system that obeys the Hamiltonian equations of motion, the existence of a Lagrangian formulation
can be proved. For a continuous (field) field systems, construction of a Lagrangian formulation is not so trivial
procedure, due to presence of spatial derivatives of the fields. For instance, to reformulate the Maxwell equations
as the second-order Lagrangian system, one needs to pass from the description in terms of electric ${\bf E}$ and
magnetic ${\bf B}$ fields to those in terms of either the three-dimensional vector potential ${\bf A}$ or the
four-dimensional vector potential $A_\mu$. The initial fields are related with the three-dimensional potential by
differential operators
\begin{eqnarray}\label{es1}
{\bf E}=-\frac{1}{c}\partial_t{\bf A}, \qquad {\bf B}=\mbox{\boldmath$\nabla$}\times{\bf A}.
\end{eqnarray}
It should be noticed that the importance of the reformulation can
not be overestimated, being the starting point for modern formulation of classical and quantum theory of electromagnetic
field.

While in some cases it requires the use of rather sophisticated methods, Lagrangian formulations have been found for
most fundamental equations of mathematical physics. One of equations which, up to present, is not included into this list, is the Schr\"odinger equation. In this work we discuss the possibility to construct a
Lagrangian formulation for the Schr\"odinger equation in the sense that we look for the second-order equation for {\it unique real field} $\phi$
that would be
{\it equivalent} to the Schr\"odinger equation\footnote{In fact, the problem has been raised already by Schr\"odinger [1].
Eq. (\ref{-48S7}) below has been tested by Schr\"odinger as a candidate for the wave function equation and then abandoned.}.
After introduction the field into the formalism, its mathematical structure becomes analogous to those of electrodynamics. In particular,
as well as ${\bf A}$ represents a potential for magnetic and electric fields, the field $\phi$ turns out to be a
potential for the wave function\footnote{We stress that it does not affects the foundations of Quantum Mechanics. We
are talking only on a similarity between the mathematical structures of the two theories.}, giving its real and imaginary parts by differentiation
\begin{eqnarray}\label{es2}
\mbox{Im}\,\Psi=\hbar\partial_t\phi, \qquad
\mbox{Re}\,\Psi=-(\frac{\hbar^2}{2m}\triangle-V)\phi.
\end{eqnarray}
So, the real field $\phi$ will be called below the wave-function potential.

The work is organized as follows. Formulation of electrodynamics in terms of the three-dimensional potential is usually
achieved by solving the pair of homogeneous equations of the Maxwell system [2, 3]. In Sect. 2 we proceed in a slightly different way, separating a pair of the Hamiltonian-type equations and replacing the other pair by appropriate initial conditions.
Hence the Maxwell system is equivalent to the Hamiltonian system which must be solved under these initial conditions. The Hamiltonian equations can be then turned out into the second order equations for the vector potential.
As it will be seen, in this setting, similarity of the electrodynamics and the quantum-mechanics formulations became transparent. In Sect. 3 we apply a similar procedure to the Schr\"odinger equation, obtaining  the Lagrangian action for the wave-function potential and demonstrate the equivalence of equation of motion for the potential to the
Schr\"odinger equation. In the Conclusion we compare the results of two sections and observe the remarkable similarities
between the formulations. They are summarized in Figure 1.

\section{Electrodynamics in terms of three-dimensional vector potential}
Moving electric charges can be described using the charge density $\rho(t, x^a)$ and the current density vector
${\bf J}(t, x^a)=\rho(t, x^a){\bf v}(t, x^a)$, where ${\bf v}$ is the velocity of a charge at $t, x^a$.
They produce the electric ${\bf E}(t, x^a)$ and the magnetic ${\bf B}(t, x^a)$ fields.
According to Maxwell, an electromagnetic field is described by six functions ${\bf E}$, ${\bf B}$ subject to eight
equations
\begin{eqnarray}\label{ch8.1}
\frac{1}{c}\partial_t{\bf E}-\mbox{\boldmath$\nabla$}\times{\bf B}=-\frac{1}{c}{\bf J},
\end{eqnarray}
\begin{eqnarray}\label{ch8.2}
\frac{1}{c}\partial_t{\bf B}+\mbox{\boldmath$\nabla$}\times{\bf E}=0,
\end{eqnarray}
\begin{eqnarray}\label{ch8.3}
\mbox{\boldmath$\nabla$}\cdot{\bf E}=\rho,
\end{eqnarray}
\begin{eqnarray}\label{ch8.4}
\mbox{\boldmath$\nabla$}\cdot{\bf B}=0.
\end{eqnarray}
We use the notation
$\mbox{\boldmath$\nabla$}$$=$$(\frac{\partial}{\partial x^1}, \frac{\partial}{\partial x^2}, \frac{\partial}{\partial x^3})$,\,    $\dot\varphi$$\equiv$$\partial_t\varphi$$=$$\frac{\partial\varphi(t, x^i)}{\partial t}$,\,
$\triangle$$=$$\frac{\partial^2}{\partial x^{i2}}$.
There are six equations of the first order with respect to time, (\ref{ch8.1}), (\ref{ch8.2}). Two more
equations (\ref{ch8.3}), (\ref{ch8.4}) do not involve the time derivative and hence represent the field analogy
of kinematic constraints. We first reduce the number of equations from eight to six. \par
\noindent
{\bf Maxwell equations as the Hamiltonian system.} A specific property of
the Maxwell system is that the constraint equations can be replaced by properly-chosen initial conditions for the
problem. Indeed, consider the following problem
\begin{eqnarray}\label{ch8.9}
\frac{1}{c}\partial_t{\bf E}-\mbox{\boldmath$\nabla$}\times{\bf B}=-\frac{1}{c}{\bf J},
\end{eqnarray}
\begin{eqnarray}\label{ch8.10}
\frac{1}{c}\partial_t{\bf B}+\mbox{\boldmath$\nabla$}\times{\bf E}=0,
\end{eqnarray}
with the initial conditions
\begin{eqnarray}\label{ch8.11}
\left.\left[\mbox{\boldmath$\nabla$}\cdot{\bf E}-\rho\right]\right|_{t=0}=0, \qquad
\left.\mbox{\boldmath$\nabla$}\cdot{\bf B}\right|_{t=0}=0.
\end{eqnarray}
This is equivalent to the problem (\ref{ch8.1})-(\ref{ch8.4}). Any solution
to (\ref{ch8.1})-(\ref{ch8.4}) satisfies the equations (\ref{ch8.9})-(\ref{ch8.11}). Conversely,
let ${\bf E}$, ${\bf B}$ be the solution to the problem (\ref{ch8.9})-(\ref{ch8.11}). Taking the divergence of
Eq. (\ref{ch8.9}), we obtain the consequence\footnote{Here we have used the continuity equation. Recall that the
charge and current densities in Maxwell equations can not be taken as arbitrary, but must obey the continuity equation
$\partial_t\rho+\mbox{\boldmath$\nabla$}\cdot{\bf J}=0$.
This follows from Eq. (\ref{ch8.1}), if we compute the divergence of both sides,
and then use the identity $\mbox{\boldmath$\nabla$}\cdot\mbox{\boldmath$\nabla$}\times{\bf B}=0$ as well as Eq. (\ref{ch8.3}).}
$\partial_t\mbox{\boldmath$\nabla$}\cdot{\bf E}+\mbox{\boldmath$\nabla$}\cdot{\bf J}=
\partial_t[\mbox{\boldmath$\nabla$}\cdot{\bf E}-\rho]=0$. The initial condition (\ref{ch8.11}) then
implies $\mbox{\boldmath$\nabla$}\cdot{\bf E}-\rho=0$, that is, Eq. (\ref{ch8.3}). In the same way, taking the divergence
of Eq. (\ref{ch8.10}) we arrive at Eq. (\ref{ch8.4}).

Considering ${\bf B}$ as the conjugate momentum for ${\bf E}$, the equations (\ref{ch8.9}), (\ref{ch8.10}) can be considered as
a Hamiltonian equations
\begin{eqnarray}\label{es3}
\dot q=\frac{\delta H(q, p)}{\delta p}, \qquad \dot p=-\frac{\delta H(q, p)}{\delta q},
\end{eqnarray}
where the Hamiltonian is
\begin{eqnarray}\label{es4}
H=\int d^3x\left[\frac{c}{2}{\bf B}\cdot\mbox{\boldmath$\nabla$}\times{\bf B}+
\frac{c}{2}{\bf E}\cdot\mbox{\boldmath$\nabla$}\times{\bf E}-
{\bf B}\cdot{\bf J}\right].
\end{eqnarray}
{\bf From an electric and magnetic fields to the vector potential.} To restore the Lagrangian formulation starting from
a given Hamiltonian equations (\ref{es3}),
it is sufficient to resolve the first equation from (\ref{es3}) with respect to $p$. Substituting the solution $p(q, \dot q)$ into
the second equation, we obtain the Lagrangian equation for the position variable $q$. Substitution of the solution
into the expression $p\dot q-H(q, p)$ gives the corresponding Lagrangian, see for example [4].

For the present case, one needs to extract ${\bf B}$ from Eq. (\ref{ch8.9}). Due to presence of spatial derivatives of ${\bf B}$,
this can not be done by algebraic methods. So, to
find a Lagrangian description, we need to reformulate the problem. To make transparent the similarity of the electrodynamics
and the quantum mechanics formalisms, we do the reformulation following to a slightly different way as compare with
the standard textbooks [2, 3].

It is convenient to unify the vectors ${\bf E}$, ${\bf B}$ into the complex field
\begin{eqnarray}\label{ch8.12}
{\bf W}\equiv{\bf B}+i{\bf E}.
\end{eqnarray}
Then Eqs. (\ref{ch8.9})-(\ref{ch8.11}) can be written in a more compact form
\begin{eqnarray}\label{ch8.13}
\left(\frac{i}{c}\partial_t+\mbox{\boldmath$\nabla$}\times\right){\bf W}=\frac{1}{c}{\bf J}, \qquad
\left.\left[\mbox{\boldmath$\nabla$}\cdot{\bf W}-i\rho\right]\right|_{t=0}=0.
\end{eqnarray}
If we look for a solution of the form
${\bf W}=(-\frac{i}{c}\partial_t+\mbox{\boldmath$\nabla$}\times){\bf A}$, the equations that appear
for ${\bf A}$ turn out to be real. They read\footnote{we use the identity $\epsilon_{cab}\epsilon_{cmn}=\delta_{am}\delta_{bn}-\delta_{an}\delta_{bm}$, then
$\mbox{\boldmath$\nabla$}\times(\mbox{\boldmath$\nabla$}\times{\bf A})=-\triangle{\bf A}+
\mbox{\boldmath$\nabla$}(\mbox{\boldmath$\nabla$}\cdot{\bf A})$.}
\begin{eqnarray}\label{ch8.14}
\frac{1}{c^2}\partial^2_t{\bf A}-\triangle{\bf A}+
\mbox{\boldmath$\nabla$}(\mbox{\boldmath$\nabla$}\cdot{\bf A})=\frac{1}{c}{\bf J},
\end{eqnarray}
\begin{eqnarray}\label{ch8.14_1}
\left.\left[\partial_t\mbox{\boldmath$\nabla$}\cdot{\bf A}+c\rho\right]\right|_{t=0}=0,
\end{eqnarray}
Hence it is consistent to take ${\bf A}$ as a real function. Thus, any real solution ${\bf A}(t, x)$ of Eq. (\ref{ch8.14})
determines a solution
\begin{eqnarray}\label{ch8.16}
{\bf B}=\mbox{\boldmath$\nabla$}\times{\bf A}, \qquad {\bf E}=-\frac{1}{c}\partial_t{\bf A}.
\end{eqnarray}
of the Maxwell equations.

Conversely, any given solution ${\bf E}$, ${\bf B}$ of the Maxwell equations can be written in the form (\ref{ch8.16}),
where ${\bf A}$ is a solution to the problem (\ref{ch8.14}), (\ref{ch8.14_1}). To see this, we construct the field
\begin{eqnarray}\label{ch8.17}
{\bf A}(t, x^i)=-c\int_0^t{\bf E}(\tau, x^i)d\tau+{\bf K}(x^i),
\end{eqnarray}
where ${\bf K}$ is any solution to the equation
\begin{eqnarray}\label{ch8.18}
\mbox{\boldmath$\nabla$}\times{\bf K}={\bf B}(0, x^i).
\end{eqnarray}
The existence of the solution ${\bf K}$ of this equation is guaranteed by the equation (\ref{ch8.11}). By direct substitution, we can verify that
the field constructed obeys the equations\footnote{we point out that the initial condition (\ref{ch8.14_1}) can be transformed back into the differential equation. It is easy to see that the problem
\begin{eqnarray}\label{ch8.27}
\frac{1}{c^2}\partial^2_t{\bf A}-\triangle{\bf A}+\mbox{\boldmath$\nabla$}(\mbox{\boldmath$\nabla$}\cdot{\bf A})-
\frac{1}{c}{\bf J}=0, \quad
\partial_t\mbox{\boldmath$\nabla$}\cdot{\bf A}+c\rho=0,
\end{eqnarray}
is equivalent to the problem (\ref{ch8.14}), (\ref{ch8.14_1}).} (\ref{ch8.16}), (\ref{ch8.14}), (\ref{ch8.14_1}).

Equations of motion (\ref{ch8.14}) for the vector potential follow from the Lagrangian action
\begin{eqnarray}\label{ch8.19}
S=\int dtd^3x\left[\frac{1}{2c^2}\partial_t{\bf A}\cdot\partial_t{\bf A}-
\frac12\mbox{\boldmath$\nabla$}\times{\bf A}\cdot\mbox{\boldmath$\nabla$}\times{\bf A}+
\frac{1}{c}{\bf A}\cdot{\bf J}\right].
\end{eqnarray}
Note that the solutions ${\bf A}(t, x^i)$ and
${\bf A}(t, x^i)+\mbox{\boldmath$\nabla$}\alpha(x^i)$, where $\alpha(x^i)$ is an arbitrary function,
determine the same ${\bf E}$, ${\bf B}$.
The action reflects this fact, being invariant under these transformation\footnote{The transformation is a reminiscence on
a gauge invariance of the four-dimensional formulation that survives in the gauge $A_0=0$.}.

In short, Maxwell equations can be reformulated as a Hamiltonian system that admits a Lagrangian formulation
in the appropriately chosen variables. \par
\noindent
{\bf Maxwell equations as a generalized Hamiltonian system.} For the later use, we mention that the free Maxwell equations
can be considered as a generalized Hamiltonian system [5, 6]. Indeed, we can rewrite (\ref{ch8.9}), (\ref{ch8.10})
with ${\bf J}=0$ in the form
\begin{eqnarray}\label{-48S10}
\partial_t{\bf E}=\{{\bf E}, H'\}', \qquad \partial_t{\bf B}=\{{\bf B}, H'\}',
\end{eqnarray}
where $H'$ is the generalized Hamiltonian
\begin{eqnarray}\label{-48S11}
H'=\int d^3x\frac{c}{2}\left[{\bf E}^2+{\bf B}^2\right],
\end{eqnarray}
and the non-canonical Poisson bracket is specified by
\begin{eqnarray}\label{-48S12}
\{E_i(t, x), B_j(t, y)\}'=-\epsilon_{iyk}\partial^x_k\delta^3(x-y).
\end{eqnarray}
In contrast to $H$, the generalized Hamiltonian $H'$ does not contain the spatial derivatives of fields.

\section{Real scalar potential of a quantum-mechanics wave function}
{\bf Schr\"odinger equation as a Hamiltonian system.} Consider the quantum mechanics of a particle subject to the potential $V(t, x^i)$. The Schr\"odinger equation for the complex wave function $\Psi(t, x^i)$
\begin{eqnarray}\label{-48S1}
i\hbar\dot\Psi=-\frac{\hbar^2}{2m}\triangle\Psi+V\Psi,
\end{eqnarray}
is equivalent to the system of two equations for two real functions (the real and imaginary parts of $\Psi$, $\Psi$$=$$\varphi$$+$$ip$). We have
\begin{eqnarray}\label{-48S2}
\hbar\dot\varphi=
-\left(\frac{\hbar^2}{2m}\triangle-V\right)p,
\end{eqnarray}
\begin{eqnarray}\label{-48S2.2}
\hbar\dot p=\left(\frac{\hbar^2}{2m}\triangle-V\right)\varphi. \quad
\end{eqnarray}
We can treat $\varphi(t, x^i)$ and $p(t, x^i)$ as coordinate and conjugate momenta of the field $\varphi$ at the space point $x^i$. Then the system has the Hamiltonian form $\dot\varphi$$=$$\{\varphi, H\}$,
$\dot p $$=$$\{p, H\}$, with the Hamiltonian being
\begin{eqnarray}\label{-48S3}
H=\frac{1}{2\hbar}\int d^3x\left(\frac{\hbar^2}{2m}(\vec\nabla\varphi\vec\nabla\varphi+
\vec\nabla p\vec\nabla p)+V(\varphi^2+p^2)\right).
\end{eqnarray}
Hence the equations (\ref{-48S2}), (\ref{-48S2.2}) arise from the variation problem with the Hamiltonian action being
\begin{eqnarray}\label{-48S4}
S_H=\int dt d^3x\left[\pi\dot\varphi-\right. \qquad \qquad \qquad \cr
\left.\frac{1}{2\hbar}\left(\frac{\hbar^2}{2m} (\vec\nabla\varphi\vec\nabla\varphi+\vec\nabla p\vec\nabla p)+V(\varphi^2+p^2)\right)\right].
\end{eqnarray}
Disregarding the boundary term, this can also be rewritten  in terms of the wave function $\Psi$ and its complex conjugate $\Psi^*$
\begin{eqnarray}\label{-48S5}
S_H=\int dt d^3x\left[\frac{i\hbar}{2}(\Psi^*\dot\Psi-
\dot\Psi^*\Psi)-
\frac{\hbar^2}{2m}\vec\nabla\Psi^*\vec\nabla\Psi-V\Psi^*\Psi\right].
\end{eqnarray}
{\bf From a wave function to the wave-function potential.} Due to the Hamiltonian nature of the Schrodinger equation, it is
natural to search for a Lagrangian formulation of the
system (\ref{-48S2}), (\ref{-48S2.2}), that is a second-order equation with respect to the time
derivative for the real
function $\varphi(t, x^i)$. As it has been already mentioned in Sect. 2, we need to solve (\ref{-48S2}) with respect to $p$
and then to substitute the result either in Eq. (\ref{-48S2.2}) or into the Hamiltonian action (\ref{-48S4}). This
leads immediately to the rather formal non-local
expression $p$$=$$\hbar(-\frac{\hbar^2}{2m}\triangle-V)^{-1}$$\partial_t\varphi$. Similarly to the case of electrodynamics,
the Schr\"odinger system cannot
be obtained starting from a Lagrangian. Nevertheless, for the case of time-independent potential $V(x^i)$,
there is a Lagrangian field theory with the property that any solution to the Schr\"odinger equation can be constructed
from a solution to this theory. To find it we look for solutions of the form
\begin{eqnarray}\label{-48S6}
\Psi=-(\frac{\hbar^2}{2m}\triangle-V)\phi+i\hbar\dot\phi,
\end{eqnarray}
where $\phi(t, x^i)$ is a {\it real} function. Inserting (\ref{-48S6}) into (\ref{-48S1}) we conclude that $\Psi$
will be a solution to the Schr\"odinger equation if $\phi$ obeys the equation
\begin{eqnarray}\label{-48S7}
\hbar^2\ddot\phi+(\frac{\hbar^2}{2m}\triangle-V)^2\phi=0,
\end{eqnarray}
which follows from the Lagrangian action
\begin{eqnarray}\label{-48S8}
S=\int dtd^3x\left[\frac{\hbar}{2}\dot\phi\dot\phi-
\frac{1}{2\hbar}(\frac{\hbar^2}{2m}\triangle-V)\phi
(\frac{\hbar^2}{2m}\triangle-V)\phi\right].
\end{eqnarray}
This can be treated as the classical theory of field $\phi$ on the given external background $V(x^i)$.  The action contains  Planck's constant as a parameter. After the rescaling $(t, x^i, \phi)$$\rightarrow$$(\hbar t, \hbar x^i, \sqrt\hbar\phi)$ it appears in the potential only, $V(\hbar x ^i)$, and thus plays the role of a coupling constant of the field $\phi$ with the background.

Similarly to electrodynamics, different potentials can lead to the same wave function. It follows from the observation that any function $\alpha(x^i)$ which obeys the equation $(\frac{\hbar^2}{2m}\triangle-V)\alpha=0$ produces a vanishing wave function
(\ref{-48S6}). So, the potentials $\phi$ and $\phi+\alpha$ produce the same wave function.

In quantum mechanics the quantity $\Psi^*\Psi$ has an interpretation as a probability density, that is the
expression $\Psi^*(t, x^i)$$\Psi(t, x^i)$$d^3x$ represents the probability of finding a particle in the volume $d^3x$
around the point $x^i$ at the instant $t$.
According to the formula (\ref{-48S6}), we write
\begin{eqnarray}\label{-48S8.1}
\Psi^*\Psi=\hbar^2(\dot\phi)^2+[(-\frac{\hbar^2}{2m}\triangle+V)\phi]^2=2\hbar E,
\end{eqnarray}
where $E$$=$$T+U$ is the energy density of the field $\varphi$. Eq. (\ref{-48S8.1}) states that the probability density is the energy density of the wave potential $\phi$.
%
%
So the preservation of probability is just an energy conservation law of the theory (\ref{-48S8}). \par
\noindent
{\bf Equivalence of equation for the wave-function potential to the Schr\"odinger equation.}
Any solution to the field theory (\ref{-48S8}) determines a solution to the Schr\"odinger equation according to Eq. (\ref{-48S6}). We should ask whether an arbitrary solution to the Schr\"odinger equation can be presented in the form (\ref{-48S6}). An affirmative answer can be obtained as follows.

Let $\Psi$$=$$\varphi+ip$ be a solution to the Schr\"odinger equation. Consider the expression (\ref{-48S6}) as an equation for determining $\phi$
\begin{eqnarray}\label{-48S9.1}
\dot\phi=\frac{1}{\hbar}p,
\end{eqnarray}
\begin{eqnarray}\label{-48S9.2}
(\frac{\hbar^2}{2m}\triangle-V)\phi=-\varphi,
\end{eqnarray}
Here the right-hand sides are known functions.
Take Eq. (\ref{-48S9.2}) at $t$$=0$, $(\frac{\hbar^2}{2m}\triangle-V)\phi$$=$$-\varphi(0, x^i)$. The elliptic equation
can be solved (at least for the analytic function $\varphi(x^i)$ [7]\,); let us denote the solution as $C(x^i)$. Then the function
\begin{eqnarray}\label{-48S9.3}
\phi(t, x^i)=\frac{1}{\hbar}\int_0^t d\tau p(\tau, x^i)+C(x^i),
\end{eqnarray}
obeys the equations (\ref{-48S9.1}), (\ref{-48S9.2}). They imply the desired result: any solution to the Shr\"odinger equation can be presented  through the field $\phi$ and its momenta according to (\ref{-48S6}). Finally,  note that Eqs. (\ref{-48S9.1}), (\ref{-48S9.2}) together with Eqs. (\ref{-48S2}), (\ref{-48S2.2}) imply that $\phi$ obeys Eq. (\ref{-48S7}). \par
\noindent
{\bf Schr\"odinger equation as the generalized Hamiltonian system.}
Let us finish this subsection with one more comment. As we have seen, treating a Schr\"odinger system  as a Hamiltonian one, it is impossible to construct the corresponding Lagrangian formulation owing to the presence of the spatial derivatives of momentum in the Hamiltonian. To avoid this problem, we can try to treat the Schr\"odinger system as a generalized Hamiltonian system. We rewrite (\ref{-48S2}) in the form
\begin{eqnarray}\label{-48S10}
\dot\varphi=\{\varphi, H'\}', \qquad
\dot p=\{p, H'\}',
\end{eqnarray}
where $H'$ is the generalized Hamiltonian
\begin{eqnarray}\label{-48S11}
H'=\int d^3x\frac{1}{2\hbar}(p^2+\varphi^2)=\int d^3x\frac{1}{2\hbar}\Psi^*\Psi,
\end{eqnarray}
and the non-canonical Poisson bracket is specified by
\begin{eqnarray}\label{-48S12}
\{\varphi, \varphi\}'=\{p, p\}'=0, \qquad \qquad \qquad \cr
\{\varphi(t, x), p(t, y)\}'=-(\frac{\hbar^2}{2m}\triangle-V)\delta^3(x-y).
\end{eqnarray}
In contrast to $H$, the Hamiltonian $H'$ does not contain the spatial derivatives of momentum.

We point out that a non-canonical bracket turns out to be a characteristic property of a singular Lagrangian theory.
This observation has been explored in the recent work [8], where we show that there is a singular Lagrangian theory
subject to second-class constraints underlying both the Schr\"odinger equation and the theory of the field $\phi$.

\section{Conclusion}
Results of the present work can be resumed as follows.\par
\noindent
1. It has been demonstrated  that the Schr\"odinger equation (\ref{-48S1}) with time-independent potential $V(x^i)$ is
equivalent to the second order equation (\ref{-48S7}) for the real function $\phi$. So, in applications one can look for solutions of the equation (\ref{-48S7}) instead the Schr\"odinger one.

The formula (\ref{-48S6}) implies
that after introduction of the field $\phi$ into the formalism, its
mathematical structure becomes analogous to that of electrodynamics. The dynamics of the
magnetic ${\bf B}$ and electric ${\bf E}$ fields is governed by first-order Maxwell equations
with respect to the time variable. Equivalently, we can use the vector potential ${\bf A}$, which
obeys the second-order equations following from the Lagrangian (\ref{ch8.19}). ${\bf A}$
represents the potential for magnetic and electric fields, generating them according
to ${\bf B}=\mbox{\boldmath$\nabla$}\times{\bf A}$,
${\bf E}=-\frac{1}{c}\partial_t{\bf A}$. Similarly to this, the field $\phi$ turns out to be a potential for the
wave function, generating its real and imaginary parts according to
Re\,$\Psi=-(\frac{\hbar^2}{2m}\triangle-V)\phi$, Im\,$\Psi=\hbar\partial_t\phi$. \par
\noindent
2. Comparing the reasonings in sections 2 and 3 we observe the remarkable similarities between mathematical
structure of electrodynamics and quantum-mechanics . They are summarized in Figure 1 on the page ~\pageref{fig_1}.
\begin{figure*}[t]\label{fig_1}
\begin{tabular}{|p{5.5cm}|p{5.5cm}|}
\hline
Electrodynamics & Quantum mechanics\\
\hline
Maxwell equations can be considered as a Hamiltonian system with the Hamiltonian $\qquad \qquad$
$\frac{c}{2}(\vec B, [\vec\nabla, \vec B])+\frac{c}{2}(\vec E, [\vec\nabla, \vec E])-
(\vec B, \vec J)$ & Schr\"odinger equations can be considered as a Hamiltonian system with the Hamiltonian $\qquad \qquad$
$\frac{\hbar}{4m}(\vec\nabla\varphi\vec\nabla\varphi+\frac{1}{2\hbar}
\vec\nabla p\vec\nabla p)+V(\varphi^2+p^2)$\\
\hline
There is a Lagrangian formulation in terms of $\vec A$ & There is a Lagrangian formulation in terms of $\phi$\\
\hline
$\vec A$ represents the potential for magnetic and electric fields, one has
$\qquad \qquad \qquad \vec B+i\vec E=\nabla\times\vec A-\frac{i}{c}\partial_t\vec A$ & $\phi$ represents the wave-function potential, one has $\qquad \qquad \qquad \qquad $
$\Psi=\varphi+ip=-(\triangle-V)\phi+i\hbar\partial_t\phi$\\
\hline
While the Maxwell equations are written in terms of $\vec B$, $\vec E$, the field $\vec E$ is the conjugate momenta for $\vec A$ but not for $\vec B$ & While the Schr\"odinger equation is written in terms of $\varphi$, $p$, the field $p$ is the conjugate momenta for $\phi$ but not for $\varphi$\\
\hline
The vector potentials $\vec A$ and $\vec A+\vec{\nabla}\alpha(x^a)$ produce the same electromagnetic field & Scalar potentials $\phi$ and $\phi+\alpha(x^a)$, where $(\frac{\hbar^2}{2m}\triangle-V)\alpha=0$ produce the same wave function\\
\hline
Free Maxwell equations form the generalized Hamiltonian system with the Hamiltonian
$\frac{c}{2}(\vec E^2+\vec B^2)$ & Schr\"odinger equation forms the generalized Hamiltonian system with the Hamiltonian
$\frac{1}{2\hbar}( p^2+\varphi^2)$\\
\hline
\end{tabular}
\caption{Real field $\phi$ as a wave-function potential.}\label{table}
\end{figure*}

\section{Acknowledgments}
The author would like to thank the Brazilian foundations CNPq (Conselho Nacional de Desenvolvimento Científico e Tecnológico - Brasil) and FAPEMIG for financial support.

\end{document}